\documentclass[onecolumn]{aa}
\usepackage{graphicx}
\begin{document}
   \title{The far-IR/radio correlation in the ISO era\thanks{Based on
   observations with the {\it Infrared Space Observatory (ISO)}, an ESA project
   with instruments funded by ESA member states (especially the PI countries:
   France, Germany, the Netherlands, and the United Kingdom) and with
   the participation of ISAS and NASA.}}

   \subtitle{The warm and cold far-IR/radio correlations}

   \author{D. Pierini
          \inst{1}
          \and
          C. C. Popescu\inst{2,3}
          \and
          R. J. Tuffs\inst{2}
          \and
          H. J. V\"olk\inst{2}
          }

   \offprints{D. Pierini}

   \institute{Max-Planck-Institut f\"ur extraterrestrische Physik,
              Giessenbachstr., D-85748 Garching\\
              \email{dpierini@mpe.mpg.de}
         \and
             Max-Planck-Institut f\"ur Kernphysik,
             Saupfercheckweg 1, D-69117 Heidelberg\\
             \email{Cristina.Popescu@mpi-hd.mpg.de}\\
             \email{Richard.Tuffs@mpi-hd.mpg.de}\\
             \email{Heinrich.Voelk@mpi-hd.mpg.de}
          \and Research Associate, The Astronomical Institute of the Romanian
              Academy, Str. Cu\c titul de Argint 5, Bucharest, Romania
             }

   \date{Received ...; accepted ...}

   \abstract{
   We present the correlation between the far-infrared (FIR) and radio
   emissions from a composite sample of 72 nearby normal galaxies observed with
   the ISOPHOT instrument on board the Infrared Space Observatory.
   The galaxies in the sample have measurements at three FIR wavelengths
   (60, 100 and 170\,${\mu}$m), which allowed a direct determination of
   the warm and cold FIR emission components.
   This is the first time that the correlation has been established
   for the total FIR luminosity, of which most is carried by the cold dust
   component predominantly emitting longwards of the spectral coverage of IRAS.
   The slope of this correlation is slightly non-linear ($1.10\pm 0.03$).
   Separate correlations between the warm and cold FIR emission components
   and the radio emission have also been derived.
   The slope of the warm FIR/radio correlation was found to be linear
   ($1.03 \pm 0.03$).
   For the cold FIR/radio correlation we found a slightly non-linear
   ($1.13 \pm 0.04$) slope.
   We qualitatively interpret the correlations in terms of star formation rate
   and find that both the FIR and radio emissions may be consistent with
   a non-linear dependence on star formation rate for galaxies not undergoing
   starburst activity.

   }

   \maketitle
%

\section{Introduction}

   One of the most surprising discoveries of the IRAS all-sky survey
   was the very tight and universal correlation between
   the spatially integrated far-infrared (FIR) and radio continuum emissions
   of late-type galaxies (de Jong et al. 1985; Helou, Soifer \& Rowan-Robinson
   1985; Wunderlich, Wielebinski \& Klein 1987).
   The standard interpretation of the so-called FIR/radio correlation
   is in terms of massive star formation activity (see Condon 1992 and 
   V\"olk \& Xu 1994 for reviews).
   This picture assumes that the FIR emission is due to dust heated by
   the massive stars which are also responsible for the radio emission:
   the ionising radiation from the massive stars powers the thermal radio
   emission, and the remnants of supernova explosions which occur at the end
   of their lives presumably accelerate the cosmic ray electrons.
   The link between the radio synchrotron and dust emissions was quantified by
   the calorimeter theory of V\"olk (1989) (see also Lisenfeld, V\"olk \& Xu
   1996a).
   Specific aspects of the FIR/radio correlation have been addressed by
   many authors (e.g. Gavazzi, Cocito \& Vettolani 1986; Cox et al. 1988;
   Devereux \& Eales 1989; Chi \& Wolfendale 1990; Condon, Anderson \& Helou
   1991; Price \& Duric 1992; Xu et al. 1994; Xu, Lisenfeld \& V\"olk 1994;
   Niklas, Klein \& Wielebinski 1995; Lisenfeld, V\"olk \& Xu 1996b; Bressan,
   Silva \& Granato 2002; Bell 2003; Groves et al. 2003).

   All these studies of the FIR/radio correlation were based on
   IRAS measurements and FIR luminosities obtained by
   using the IRAS flux densities at 60 and 100\,$\rm {\mu} m$.
   However the FIR spectral energy distribution (SED) of
   normal\footnote{By ``normal'' we loosely refer to galaxies which are not
   dominated by an active nucleus and whose current star formation rates
   would be sustainable for a substantial fraction of a Hubble time.} galaxies
   typically peaks longwards of 100\,$\rm {\mu} m$, beyond the spectral grasp
   of IRAS, as predicted from submillimeter (submm) observations by
   Chini et al. (1986).
   With the advent of the Infrared Space Observatory (ISO)
   (Kessler et al. 1996), in particular of the ISOPHOT instrument
   (Lemke et al. 1996) on board ISO, it became routinely possible to measure
   the peak of the FIR emission, also for fainter objects not detected by IRAS.
   It was found that, especially for quiescent galaxies, most of
   the FIR luminosity is carried by a cold dust component (Popescu et al. 2002)
   whose luminosity is not well predicted by the IRAS measurements.
   In fact the luminosity correction factors\footnote{These correction factors
   are the multiplicative factors by which the total FIR luminosity
   (40-1000\,$\rm {\mu} m$), as derived from ISO measurements,
   differs from the IRAS FIR luminosity, derived from the formula
   (Helou et al. 1988): $f_{40-120}[{\rm W m^{-2}}]=1.26\times 10^{-14}(2.58f_{60{\rm \mu m}}+f_{100{\rm \mu m}})$, with $f_{60{\rm \mu m}}$
   and $f_{100{\rm \mu m}}$ expressed in Jy.} for cold dust do not correlate
   with Hubble type and exhibit a huge scatter, with mean values ranging from
   1 to 3 (Popescu \& Tuffs 2002b).
   In view of this, there is a clear need to redefine the FIR/radio correlation
   in terms of the total FIR--submm luminosity, as derived from
   ISO measurements.

   Another result to emerge from ISO is the confirmation that
   the FIR SEDs of normal galaxies require a warm and a cold dust
   emission component to be fitted (for a review see Tuffs \& Popescu 2002).
   Although the concept of warm and cold emission components is as old as IRAS 
   (de Jong et al. 1984), it only became possible to directly measure
   and separate these components with ISO, due to its spectral grasp
   and multi filter coverage of the FIR regime.
   The warm component can be identified with locally heated dust
   in HII regions\footnote{In the context of dust emission we use the term
   HII regions to denote not only the optical emitting HII regions
   within the star forming complexes, but also their immediate vicinity
   where grains can also be strongly heated by the massive stars.}
   and the cold component with dust distributed in the general
   diffuse interstellar medium and heated by a combination of
   non-ionising ultraviolet (UV) and optical/near-IR photons.
   This interpretation is consistent with what has been seen
   in the ISOPHOT maps of nearby galaxies (e.g. Haas et al 1998;
   Hippelein et al. 2003).
   It is also consistent with models of the UV/optical/FIR/submm SEDs
   in normal galaxies, which self-consistently calculate
   a continuous distribution of dust temperatures based on
   radiative-transfer calculations (Silva et al. 1998; Popescu et al. 2000;
   Misiriotis et al. 2001).
   Thus ISO measurements give us the first direct opportunity of defining
   the FIR/radio correlation separately for the warm and cold dust components.
   In the past this could only be indirectly attempted by constraining
   the warm dust component to be proportional to the H$\alpha$ emission,
   as done by Xu, Lisenfeld \& V\"olk (1994).
   These authors established separate relations for the warm FIR/radio
   correlation and the cold FIR/radio correlation.
   However, the cold FIR luminosity derived in this study was based on
   IRAS data, and misses the bulk of the true cold dust luminosity.
   Other authors have used different operational recipes with different
   physical interpretations to derive warm and cold FIR emission components
   from IRAS data for comparison with radio data (e.g. Fitt, Alexander \& Cox
   1988; Wunderlich \& Klein 1991; Condon et al. 1991; Beck \& Golla 1988;
   Hoernes, Berkhuijsen \& Xu 1998), but again these studies missed the bulk of
   the cold dust emission. 

   In this paper we redefine the FIR/radio correlation using for the first time
   the total FIR luminosity (as measured with ISO) for a statistical sample.
   Plots of the FIR/radio correlation using the new ISO measurements
   were also given in Popescu et al. (2002), but only for a very small sample,
   and therefore no attempt was made to quantify the linearity of
   the correlation there.
   The goal of this paper is to provide a statistical analysis of
   the correlation and also to derive the slope of the correlation separately
   for the warm and cold dust components.
   Finally the new results are discussed in the context of predictions
   of theoretical models on photon heating of dust in normal
   star-forming galaxies.

   The paper is organised as follows.
   In Sect.~2 we describe the sample selection and the analysis of
   the radio data.
   In Sect.~3 we present the results for the total FIR/radio correlation,
   as well as for the warm and cold FIR/radio correlations.
   Each of these correlations is qualitatively analysed in terms
   of its dependence on the star formation rate in Sect.~4.
   In Sect.~5 we discuss scenarios which can simultaneously account for
   both the warm and cold FIR/radio correlations.
   A summary is given in Sect.~6.
%

\section{The ISO sample: selection and data analysis}

   The sample used in this study is a combination of a very deep,
   optically selected sample of normal, relatively quiescent galaxies
   together with a shallow, FIR selected sample of relatively bright galaxies.
   This provides us with a high dynamic range in FIR luminosities as well as
   good statistics.
   For this purpose we used the ISOPHOT Virgo Cluster Deep Survey
   (Tuffs et al. 2002a,b), which is the deepest survey (both in luminosity
   and surface brightness terms) of normal galaxies performed by ISOPHOT,
   and the ISOPHOT 170\,$\rm \mu m$ Serendipity Survey (Stickel et al. 2000).
   All galaxies in the combined sample have had their integrated flux densities
   measured by ISOPHOT at 170\,$\rm \mu m$ (the Virgo galaxies also
   in the ISOPHOT bands centered at 60 and 100\,$\rm \mu m$).
   Most galaxies from the Virgo sample are freshly falling in from the field
   (Tully \& Shaya 1984; Tuffs et al. 2002a,b) and therefore we consider it
   legitimate to combine it with the ISOPHOT 170\,$\rm \mu m$
   Serendipity Survey sample.
   Furthermore there is no evidence for an effect of the cluster environment
   on the FIR properties of the galaxies from the ISOPHOT Virgo Cluster
   Deep Sample (Popescu et al. 2002; Popescu \& Tuffs 2002b).
   This is to be expected since the bulk of the FIR emission arises primarily
   from the inner regions of the galactic disks.
   The radio emission may be more prone to alterations due to
   environmental effects, as we will discuss in Sect.~3.
   However we will show there that, in the case of our sample,
   these effects are only of second order.
   Thus, in this paper we regard the main distinction between Virgo
   and Serendipity Survey galaxies to be the quiescent nature of the former,
   rather than their association with the cluster environment.
   So, put together, the combined sample will provide us with larger statistics
   and a good dynamic range in star formation activity.

   In total 72 galaxies were selected, as described below.
   Of these, 20 galaxies show some evidence for LINER/Seyfert activity
   or are pair/multiple/merger systems. 

\subsection{The VIRGO subsample}

   The ISOPHOT Virgo Cluster Deep Sample consists of 63 galaxies later than S0
   and brighter than $B_\mathrm{T} = 16.8$, selected from the Virgo Cluster
   Catalogue (VCC) of Binggeli, Sandage \& Tammann (1985).
   Tuffs et al. (2002a,b) have presented deep diffraction-limited FIR strip 
   maps
   of this sample, obtained with ISOPHOT at the central wavelengths
   60, 100 and 170\,$\rm \mu m$\footnote{A subsample of the ISOPHOT
   Virgo Cluster Deep Sample has been also observed with
   the Long Wavelength Spectrometer at the 158\,$\rm \mu m$
   [CII] fine-structure gas cooling line (Leech et al. 1999) and analysed by
   Pierini and collaborators (1999, 2001, 2003).}.
   Details of observations, data reduction, estimate of
   the calibration uncertainty and procedure of extraction of photometry
   are given in Tuffs et al. (2002a,b).
   As discussed by these authors, the 63 galaxies represent a sample
   of nearby normal (i.e. non-AGN dominated and non-starburst)
   late-type galaxies.
   Nevertheless a few objects have also some mild LINER/Seyfert activity.

   Out of this sample, 38 galaxies (i.e. 63.5 per cent of the total)
   were detected at all three wavelengths.
   Most of them were discovered to contain a cold dust emission component
   which could not have been recognised by IRAS, irrespective of
   the morphological classification of individual objects
   (Popescu et al. 2002).
   The ISOPHOT FIR SEDs of these galaxies were fitted with a combination
   of two modified blackbody functions, physically identified with
   the emissions from a localised warm dust component, with a fixed temperature
   of 47\,$\rm K$, and a diffuse cold dust component (see Popescu et al.).
   The cold dust temperatures were found to span a broad range,
   with a median of 18\,$\rm K$.
   The physical justification for fixing the temperature of the warm component
   is based on the modelling work of Popescu et al. (2000)
   and Misiriotis et al. (2001).
   The latter papers showed that $\sim 20\%$ of the integrated FIR emission
   at 60\,${\mu}$m comes from stochastically heated grains
   in the diffuse interstellar medium (ISM), $\sim 20\%$ from optically heated
   grains in the diffuse ISM, and  $\sim 60\%$ from HII regions.
   So it is a good physical approximation to consider that most of the emission
   at $60\,{\mu}$m comes from grains within the HII regions (which do not emit
   stochastically).
   Although there is a variation in grain temperature between HII regions
   in any given galaxy (Peeters et al. 2002) (due to changes in size, geometry
   and exciting stars), these are quite moderate (in comparison to
   the systematic difference in temperature between grains in HII regions
   and grains in the diffuse ISM).
   Thus, if we average over the ensemble of HII regions in each galaxy
   we would expect the ensemble of HII regions to have virtually identical SEDs
   from one galaxy to the next.
   
   Fourteen of the 38 galaxies with both warm and cold dust
   emission components have also been detected in the radio continuum
   at 1.4 GHz, thanks to the recent NRAO/VLA Sky Survey (NVSS)
   (Condon et al. 1998).
   For these objects we performed aperture photometry to extract
   integrated flux densities from the NVSS maps (see Appendix A).
   The 14 VCC galaxies with FIR and radio counterparts are considered
   in the present study.
   They constitute the so-called ``VIRGO'' subsample.
   \begin{table}
      \caption[]{Galaxy parameters for the VIRGO subsample}
         \label{Tab1}
     $$ 
         \begin{array}{p{0.07\linewidth}lccccccc}
            \hline
            \noalign{\smallskip}
            Den. & {\mathrm{Hubble~type}} & log~L_{\mathrm{1.4 GHz}} &
            log~L_{\mathrm{FIR}} & log~{L^{\mathrm{warm}}_{\mathrm{FIR}}} & 
log~{L^{\mathrm{cold}}_{\mathrm{FIR}}} & 
log~L_{\mathrm{40-120}} & \mathrm{notes} \\
            VCC & & {[\mathrm{W~Hz^{-1}~h^{-2}}]} & {[\mathrm{W~h^{-2}}]} & {[\mathrm{W~h^{-2}}]} & 
{[\mathrm{W~h^{-2}]}} & {[\mathrm{W~h^{-2}}]} & \\
            \noalign{\smallskip}
            \hline
66 & {\mathrm{SBc}} & 20.87 \pm 0.21 & 35.71 & 35.22 & 35.53 & 35.36 & \\
92 & {\mathrm{Sb:}} & 21.18 \pm 0.16 & 36.19 & 35.52 & 36.09 & 35.67 & {\mathrm{LINER^{ab}}} \\
152 & {\mathrm{Scd}} & 20.53 \pm 0.23 & 35.60 & 35.02 & 35.46 & 35.31 & \\
460 & {\mathrm{Sa~pec}} & 20.54 \pm 0.23 & 35.78 & 35.21 & 35.64 & 35.55 & {\mathrm{LINER^c}} \\
873 & {\mathrm{Sc}} & 20.65 \pm 0.23 & 35.90 & 35.41 & 35.74 & 35.57 & \\
971 & {\mathrm{Sd}} & 19.97 \pm 0.12 & 34.80 & 34.22 & 34.67 & 34.50 & \\
1002 & {\mathrm{SBc}} & 20.06 \pm 0.15 & 35.37 & 34.67 & 35.27 & 35.06 & \\
1043 & {\mathrm{Sb(tides)}} & 21.08 \pm 0.18 & 35.83 & 35.16 & 35.72 & 35.44 & {\mathrm{LINER^d}} \\
1110 & {\mathrm{Sab~pec}} & 20.34 \pm 0.21 & 35.63 & 35.08 & 35.48 & 35.24 & {\mathrm{LINER^e}} \\
1379 & {\mathrm{SBc}} & 20.10 \pm 0.16 & 35.37 & 34.81 & 35.24 & 35.09 & \\
1554 & {\mathrm{Sm}} & 21.33 \pm 0.13 & 35.85 & 35.55 & 35.55 & 35.67 & \\
1575 & {\mathrm{SBm~pec}} & 19.96 \pm 0.12 & 35.21 & 34.84 & 34.96 & 34.99 & \\
1690 & {\mathrm{Sab}} & 21.13 \pm 0.17 & 36.12 & 35.59 & 35.97 & 35.81 & {\mathrm{LINER,~Sy^{dfg}}} \\
1727 & {\mathrm{Sab}} & 21.26 \pm 0.15 & 36.07 & 35.43 & 35.95 & 35.66 & {\mathrm{LINER,~Sy1.9^{df}}} \\
            \noalign{\smallskip}
         \end{array}
     $$
\begin{list}{}{}
\item[$^{\mathrm{a}}$] Rauscher (1995);
$^{\mathrm{b}}$ Barth et al. (1998);
$^{\mathrm{c}}$ Ho et al. (1995);
$^{\mathrm{d}}$ Ho et al. (1997);
$^{\mathrm{e}}$ Gonzalez-Delgado et al. (1997);
$^{\mathrm{f}}$ Stauffer (1982);
$^{\mathrm{g}}$ Keel (1983).
\end{list}
   \end{table}

   Table 1 reproduces the parameters relevant to this investigation
   for the individual VIRGO objects as follows: \newline
   Col. 1: VCC denomination; \newline
   Col. 2: morphological type (from the VCC, as in Tuffs et al. 2002a,b);
   \newline
   Col. 3: total radio continuum luminosity at 1.4 GHz
   ($L_{{\mathrm{1.4 GHz}}}$) with its error (see Appendix A),
   in decimal logarithmic units; \newline
   Col. 4: total FIR continuum luminosity ($L_{\mathrm{FIR}}$),
   in decimal logarithmic units.
   The uncertainty in this luminosity is $\sim15$ per cent (random) for 
   all the sample objects. \newline
   Col. 5,6: FIR continuum luminosities from the warm and cold
   dust emission components ($L_{\mathrm{FIR}}^{\mathrm{warm}}$ and
   $L_{\mathrm{FIR}}^{\mathrm{cold}}$), respectively
   (from Popescu et al. 2002), in decimal logarithmic units.
   We adopt a conservative value of 20 per cent (random) for the uncertainty
   of each of these luminosities for all the sample objects. \newline
   Col. 7: $L_{40-120}$, i.e., the FIR luminosity that IRAS would have
   derived based on the Helou et al. (1988) formula:
   $f_{40-120}[{\rm W~m^{-2}}]=1.26\times
   10^{-14}(2.58f_{60{\rm \mu m}}+f_{100{\rm \mu m}})$, where
   $f_{60{\rm \mu m}}$ and $f_{100{\rm \mu m}}$ are expressed in Jy.
   The luminosity is given in decimal logarithmic units.
   We adopt a conservative value of 30 per cent for the uncertainty of this
   luminosity for all sample objects.\newline
   Col. 8: individual notes.

   Luminosities are determined by assuming that all the galaxies
   have the same distance of $11.5~h^{-1}$ Mpc (Binggeli, Popescu
   \& Tammann 1993), where $h$ is the Hubble constant in units of
   $\rm 100~km~s^{-1}~Mpc^{-1}$.

\subsection{The ISOPHOT Serendipity Survey subsample}

   A sample of 115 nearby galaxies was presented by Stickel et al. (2000)
   as a first result of the ``ISOPHOT 170\,$\rm \mu m$ Serendipity Survey''
   (ISS).
   We refer the reader to these authors for details about observations,
   data reduction, estimate of the calibration uncertainty, procedure
   of extraction of photometry and optical/FIR identification.

   As recognised by Stickel et al., several of their objects
   have Seyfert/LINER activity and/or are pair/multiple/merging systems.
   
   Not all galaxies from the ISS catalogue could be used in the present
   study. Out of the 115 ISS galaxies we first rejected 3 not having IRAS 
   detections at both
   60 and 100\,${\mu}$m and 13 galaxies which either have no morphological 
   information or are classified as S0, elliptical or peculiar. Of the 
   remaining 99 galaxies we omitted 3 already included in the VIRGO subsample. 
   Furthermore, 26 galaxies of the remaining 96 did not have a radio 
   counterpart or were confused in the NVSS survey at 1.4 GHz. Radio 
   counterparts are 
   associated with optically catalogued galaxies when the distance of the 
   peak radio surface brightness of the potential radio identification from 
   the optical position of the target galaxy is within the NVSS resolution.
   Furthermore, one galaxy was dropped because it has an inclination greater 
   than $\rm \sim 70^o$ and an apparent major axis
   longer than $\rm \sim 4^{\prime}$. As discussed in Appendix B, 
   such galaxies may have flux densities overestimated by a factor of 2.    
   Finally, only galaxies with heliocentric velocities higher than
   1500\,$\rm km~s^{-1}$ are selected.
   The reason is as follows.
   For most of the galaxies in the ISS catalogue only kinematical distances 
   are available from individual redshifts, as listed by Stickel et al. (2000).
   The assumption of a free Hubble Flow does not apply to
   the relatively nearby galaxies.
   Hence 11 galaxies which are not members of the Virgo Cluster
   and have heliocentric velocities lower than 1500\,$\rm km~s^{-1}$
   were dropped from the sample.

   Hereafter we refer to the final 58 galaxies selected
   from the ISS catalogue as the ISS subsample.
   For all galaxies in the ISS subsample we determined {\it new} warm and 
   cold dust temperatures from the composite IRAS and ISOPHOT data listed by 
   Stickel et al. using the method described in Popescu et al. (2002).
   For this purpose we first applied ISO/IRAS corrections for the original 
   IRAS flux densities at 60 and 100\,$\rm \mu m$ and ISO/DIRBE corrections 
   for the ISOPHOT flux densities at 170\,$\rm \mu m$, both corrections being 
   taken from  Tuffs et al. (2002a,b). The SEDs of all these objects required 
   both warm and cold dust emission components to be fitted.
   \begin{table}
      \caption[]{Galaxy parameters for the ISS subsample}
         \label{Tab2}
     $$ 
         \begin{array}{p{0.18\linewidth}lcccccccc}
            \hline
            \noalign{\smallskip}
            Den. & {\mathrm{Hubble~type}} & log~L_{\mathrm{1.4~GHz}} &
log~L_{\mathrm{FIR}} & log~{L^{\mathrm{warm}}_{\mathrm{FIR}}} &
log~{L^{\mathrm{cold}}_{\mathrm{FIR}}} & log~L_{\mathrm{40-120}} &
{\mathrm{Dist.}} & \mathrm{notes} \\
            NGC/UGC or other & & {[\mathrm{W~Hz^{-1}}]} & {[\mathrm{W~h^{-2}}]} &
{[\mathrm{W~h^{-2}}]} & {[\mathrm{W~h^{-2}}]} & {[\mathrm{W~ h^{-2}}]} &
{[\mathrm{Mpc~h^{-1}}]} & \\
            \noalign{\smallskip}
            \hline
NGC\,7821 & {\mathrm{Scd~pec~sp}} & 22.37 \pm 0.23 & 37.19 & 36.77 & 36.98 & 36.99 & 73.39 & \\
NGC\,157 & {\mathrm{SAB(rs)bc}} & 21.83 \pm 0.11 & 36.88 & 36.28 & 36.75 & 36.47 & 16.68 & \\
MCG\,-05-03-020 & {\mathrm{SAB(r)c:}} & 21.29 \pm 0.11 & 36.73 & 35.91 & 36.66 & 36.26 & 56.09 & {\mathrm{AGN?^{ab}}} \\
UGC\,816 & {\mathrm{Sc}} & 22.35 \pm 0.19 & 36.90 & 36.53 & 36.65 & 36.72 & 51.88 & \\
NGC\,477 & {\mathrm{SAB(s)c}} & 21.45 \pm 0.13 & 36.72 & 36.28 & 36.52 & 36.42 & 58.76 & \\
NGC\,520 & & 22.07 \pm 0.11 & 37.17 & 36.93 & 36.81 & 36.97 & 23.37 & {\mathrm{GPair?^c}} \\
NGC\,549 & {\mathrm{(R^{\prime})SB(s)0/a}} & 21.64 \pm 0.18 & 36.53 & 35.98 & 36.39 & 36.16 & 61.76 & \\
MCG\,-05-05-007 & {\mathrm{Sbc}} & 21.41 \pm 0.13 & 36.61 & 35.84 & 36.53 & 36.09 & 59.04 & \\
UGC\,1560 & {\mathrm{(R^{\prime})SB(s)b:}} & 22.08 \pm 0.21 & 37.00 & 36.64 & 36.76 & 36.83 & 84.91 & \\
UGC\,2238 & {\mathrm{Im?}} & 22.58 \pm 0.18 & 37.46 & 37.05 & 37.25 & 37.30 & 64.36 & {\mathrm{LINER^d}} \\
NGC\,1087 & {\mathrm{SAB(rs)c}} & 21.60 \pm 0.13 & 36.69 & 36.15 & 36.54 & 36.26 & 15.19 & \\
MCG\,-03-12-002 & & 22.66 \pm 0.22 & 37.53 & 37.03 & 37.37 & 37.45 & 95.42 & {\mathrm{GPair^{e}}} \\
UGC\,3066 & {\mathrm{SAB(r)d}} & 21.57 \pm 0.22 & 36.55 & 35.95 & 36.42 & 36.20 & 46.39 & \\
NGC\,1614 & {\mathrm{SB(s)c~pec}} & 22.57 \pm 0.13 & 37.59 & 37.03 & 37.44 & 37.55 & 47.78 & {\mathrm{Sy2^{fg}}} \\
NGC\,1667 & {\mathrm{SAB(r)c}} & 22.28 \pm 0.18 & 37.15 & 36.73 & 36.95 & 36.90 & 45.46 & {\mathrm{Sy2^{h}}} \\
MCG\,-03-13-051 & {\mathrm{(R^{\prime})SB(s)b}} & 21.75 \pm 0.20 & 36.89 & 36.28 & 36.77 & 36.49 & 66.40 & \\
NGC\,2958 & {\mathrm{S(r)bc}} & 21.82 \pm 0.21 & 36.85 & 36.39 & 36.67 & 36.56 & 66.63 & \\
NGC\,3183 & {\mathrm{SB(s)bc}} & 21.69 \pm 0.22 & 36.62 & 36.09 & 36.47 & 36.41 & 30.88 & \\
MCG\,-06-23-029 & {\mathrm{Sbc:~sp}} & 21.35 \pm 0.23 & 36.26 & 35.90 & 36.02 & 36.06 & 31.22 & \\
NGC\,4222 & {\mathrm{Sc}} & 19.88 \pm 0.10 & 35.35 & 34.77 & 35.21 & 35.00 & 11.50 & \\
NGC\,4383 & {\mathrm{Sa?~pec}} & 20.80 \pm 0.22 & 35.99 & 35.72 & 35.65 & 35.77 & 11.50 & \\
NGC\,4639 & {\mathrm{SAB(rs)bc}} & 20.19 \pm 0.18 & 35.42 & 34.85 & 35.28 & 35.16 & 11.50 & {\mathrm{Sy1.8^{h}}} \\
MCG\,-05-31-035 & {\mathrm{SB(rs)d}} & 20.81 \pm 0.21 & 35.81 & 35.37 & 35.61 & 35.54 & 22.16 & \\
NGC\,4981 & {\mathrm{SAB(r)bc}} & 20.92 \pm 0.23 & 36.15 & 35.59 & 36.01 & 35.83 & 16.86 & {\mathrm{LINER^{i}}} \\
MCG\,-05-31-039 & {\mathrm{SB(s)d:~sp}} & 20.72 \pm 0.17 & 35.89 & 35.38 & 35.73 & 35.58 & 23.81 & \\
IC\,4221 & {\mathrm{SA(r)c~pec?}} & 20.87 \pm 0.16 & 35.98 & 35.50 & 35.81 & 35.70 & 28.95 & \\
NGC\,5085 & {\mathrm{SA(s)c}} & 21.24 \pm 0.22 & 36.38 & 35.71 & 36.27 & 35.98 & 19.56 & \\
NGC\,5292 & {\mathrm{(R^{\prime})SA(rs)ab}} & 21.34 \pm 0.18 & 36.41 & 35.75 & 36.30 & 36.05 & 44.66 & \\
UGC\,8739 & {\mathrm{SB?}} & 22.50 \pm 0.16 & 37.25 & 36.86 & 37.02 & 37.02 & 50.95 & \\
MCG\,-04-33-013 & {\mathrm{Sa:~pec~sp}} & 21.52 \pm 0.23 & 36.55 & 35.91 & 36.44 & 36.19 & 34.55 & \\
NGC\,5468 & {\mathrm{SAB(rs)cd}} & 21.65 \pm 0.21 & 36.52 & 36.11 & 36.31 & 36.32 & 28.45 & \\
NGC\,5504 & {\mathrm{SAB(s)bc}} & 21.77 \pm 0.23 & 36.65 & 36.19 & 36.47 & 36.40 & 52.47 & \\
NGC\,5533 & {\mathrm{SA(rs)ab}} & 21.59 \pm 0.23 & 36.39 & 35.62 & 36.31 & 36.12 & 38.66 & \\
NGC\,5604 & {\mathrm{Sa~pec?}} & 21.37 \pm 0.23 & 36.41 & 35.96 & 36.23 & 36.15 & 27.48 & \\
UGC\,9483 & {\mathrm{S}} & 20.72 \pm 0.22 & 35.79 & 35.28 & 35.63 & 35.53 & 16.36 & \\
NGC\,5757 & {\mathrm{(R^{\prime})SB(r)b}} & 21.54 \pm 0.22 & 36.65 & 36.29 &
            36.40 & 36.38 & 26.30 & \\
NGC\,5899 & {\mathrm{SAB(rs)c}} & 21.62 \pm 0.21 & 36.65 & 36.06 & 36.52 & 36.27 & 25.62 & {\mathrm{Sy2^{j}}} \\
NGC\,5956 & {\mathrm{Scd?}} & 20.70 \pm 0.19 & 35.47 & 34.80 & 35.36 & 35.24 & 18.98 & \\
NGC\,5961 & {\mathrm{S?}} & 21.27 \pm 0.22 & 35.74 & 35.35 & 35.51 & 35.51 & 17.66 & \\
NGC\,5985 & {\mathrm{SAB(r)b}} & 21.25 \pm 0.23 & 36.41 & 35.48 & 36.36 & 35.93 & 25.17 & {\mathrm{LINER,~Sy1^{k}}} \\
NGC\,6052 & & 22.42 \pm 0.15 & 37.04 & 36.79 & 36.67 & 36.87 & 44.99 & {\mathrm{GPair^{d}}} \\
UGC\,10331 & {\mathrm{S~pec}} & 21.33 \pm 0.16 & 36.51 & 36.10 & 36.30 & 36.22 & 44.71 & \\
NGC\,6120 & {\mathrm{pec}} & 22.57 \pm 0.23 & 37.47 & 37.06 & 37.26 & 37.33 & 91.70 & \\
UGC\,10524 & {\mathrm{(R^{\prime})SB(s)ab}} & 22.25 \pm 0.23 & 37.23 & 36.85 & 37.00 & 37.11 & 75.98 & \\
NGC\,6381 & {\mathrm{SA(s)c?}} & 21.22 \pm 0.20 & 36.20 & 35.80 & 35.99 & 35.98 & 32.65 & \\
NGC\,6478 & {\mathrm{SAc:}} & 22.15 \pm 0.23 & 37.10 & 36.48 & 36.99 & 36.75 & 67.75 & \\
NGC\,6574 & {\mathrm{SAB(rs)bc:}} & 21.86 \pm 0.16 & 36.89 & 36.55 & 36.63 & 36.70 & 22.52 & {\mathrm{Sy^{h}}} \\
NGC\,6585 & {\mathrm{S?}} & 21.45 \pm 0.23 & 36.39 & 35.93 & 36.21 & 36.15 & 26.35 & \\
NGC\,6667 & {\mathrm{SABab?~pec}} & 21.28 \pm 0.23 & 36.42 & 35.91 & 36.26 & 36.22 & 25.82 & \\
NGC\,6764 & {\mathrm{SB(s)bc}} & 21.97 \pm 0.15 & 36.60 & 36.32 & 36.29 & 36.42 & 24.16 & {\mathrm{LINER,~Sy2^{l}}} \\
NGC\,6796 & {\mathrm{Sbc:~sp}} & 21.30 \pm 0.23 & 36.36 & 35.81 & 36.22 & 36.20 & 21.89 & \\
NGC\,6911 & {\mathrm{SBb:}} & 21.53 \pm 0.22 & 36.60 & 36.14 & 36.41 & 36.28 & 25.01 & \\
NGC\,6928 & {\mathrm{SB(s)ab}} & 21.66 \pm 0.22 & 36.83 & 36.26 & 36.69 & 36.52 & 47.07 & \\
NGC\,7042 & {\mathrm{Sb}} & 21.78 \pm 0.23 & 36.57 & 35.97 & 36.44 & 36.30 & 50.82 & \\
NGC\,7172 & {\mathrm{Sa~pec~sp}} & 21.56 \pm 0.22 & 36.59 & 36.24 & 36.34 & 36.40 & 26.03 & {\mathrm{Sy2^{m}}} \\
IC\,1438 & {\mathrm{(R_1 R_2^{\prime})SAB(r)a}} & 20.66 \pm 0.11 & 36.00 & 35.42 & 35.86 & 35.65 & 26.16 & \\
NGC\,7541 & {\mathrm{SB(rs)bc~pec:}} & 22.22 \pm 0.11 & 37.22 & 36.86 & 36.97 & 36.98 & 26.78 & \\
NGC\,7755 & {\mathrm{SB(r)bc}} & 21.54 \pm 0.23 & 36.53 & 35.99 & 36.39 & 36.26 & 29.63 & \\
            \hline
         \end{array}
     $$
\begin{list}{}{}
\item[$^{\mathrm{a}}$] Boller et al. (1997);
$^{\mathrm{b}}$ Page et al. (2001);
$^{\mathrm{c}}$ Hibbard \& van Gorkom (1996);
$^{\mathrm{d}}$ Smith et al. (1996);
$^{\mathrm{e}}$ Vorontsov-Velyaminov \& Arhipova (1968);
$^{\mathrm{f}}$ Rodriguez Espinosa et al. (1987);
$^{\mathrm{g}}$ Risaliti et al. (2000);
$^{\mathrm{h}}$ Gonzalez-Delgado et al. (1997);
$^{\mathrm{i}}$ Bonatto et al. (1989);
$^{\mathrm{j}}$ Stauffer (1982);
$^{\mathrm{k}}$ Ho et al. (1995);
$^{\mathrm{l}}$ Alonso-Herrero et al. (2000);
$^{\mathrm{m}}$ Sharples et al. (1984).
\end{list}
   \end{table}

   We assigned distances to the members of the ISS subsample as follows:
   We identify 3 galaxies from the ISS subsample as members of the Virgo 
   cluster (and not in the ISOPHOT Virgo Cluster Deep sample) and, thus, we 
   assign them to be at the common distance of $\rm 11.5~h^{-1}~Mpc$ adopted 
   for the VIRGO subsample (cf. Sect. 2.1).
   For the other galaxies of the ISS subsample
   we have no clue whether they belong to the field
   or to a group/cluster, though we know if they are mergers or not.
   We determine kinematical distances, corrected to the Galactic Standard
   of Rest with the NED Velocity Correction Calculator, independent of
   whether they are isolated or not and whether they are mergers or not.

   Table 2 reproduces the parameters relevant to this investigation
   for the individual ISS objects as follows: \newline
   Col. 1: galaxy denomination (from Stickel et al. 2000); \newline
   Col. 2: morphological type (from NED); \newline
   Col. 3: total radio continuum luminosity at 1.4 GHz
   ($L_{{\mathrm{1.4 GHz}}}$) with its error (see Appendix A),
   in decimal logarithmic units; \newline
   Col. 4: total FIR continuum luminosity ($L_{\mathrm{FIR}}$),
   in decimal logarithmic units.
   The uncertainty in this luminosity is $\sim15$ per cent (random) for 
   all the sample objects. \newline
   Col. 5,6: FIR continuum luminosities from the warm and cold dust
   emission components ($L_{\mathrm{FIR}}^{\mathrm{warm}}$ and
   $L_{\mathrm{FIR}}^{\mathrm{cold}}$), respectively (this work),
   in decimal logarithmic units.
   We adopt a conservative value of 20 per cent (random) for the uncertainty
   of each of these luminosities for all the sample objects. \newline
   Col. 7: the IRAS FIR luminosity $L_{40-120}$ (calculated as in Table 1),
   in decimal logarithmic units.
   We adopt a conservative value of 30 per cent for the uncertainty
   of this luminosity for all the sample objects. \newline
   Col. 8: galaxy distance; \newline
   Col. 9: individual notes.
%

\section{Results}

\subsection{The total FIR/radio correlation}

   The relationship between the total (warm $+$ cold) FIR emission,
   $L_{\rm FIR}$, and the radio continuum emission at 1.4 GHz
   for the 72 sample galaxies is reproduced in Fig. 1.
   Here the solid line reproduces the linear fit to the data (in log--log)
   of equation:
   \[ \log~L_\mathrm{1.4 GHz} =  m(\pm \epsilon(m))~\log~L_\mathrm{FIR}+c (\pm \epsilon(c))~~\rm W~Hz^{-1}~h^{-2}\,, \]
   where the parameters of this fit are given in Table 3, together with
   its reduced $\chi^2$ and the dispersion around the fitted relation.
   In obtaining the fits we adopted the algorithm executing
   a bivariate least-squares fit, taking into account the uncertainties
   in both the x- and y-variables at the same time, as implemented in the task
   ``FITEXY'' of the Numerical Recipes (Press et al. 1992).
   \begin{figure}[htb]
   \centering
   \includegraphics[width=12cm]{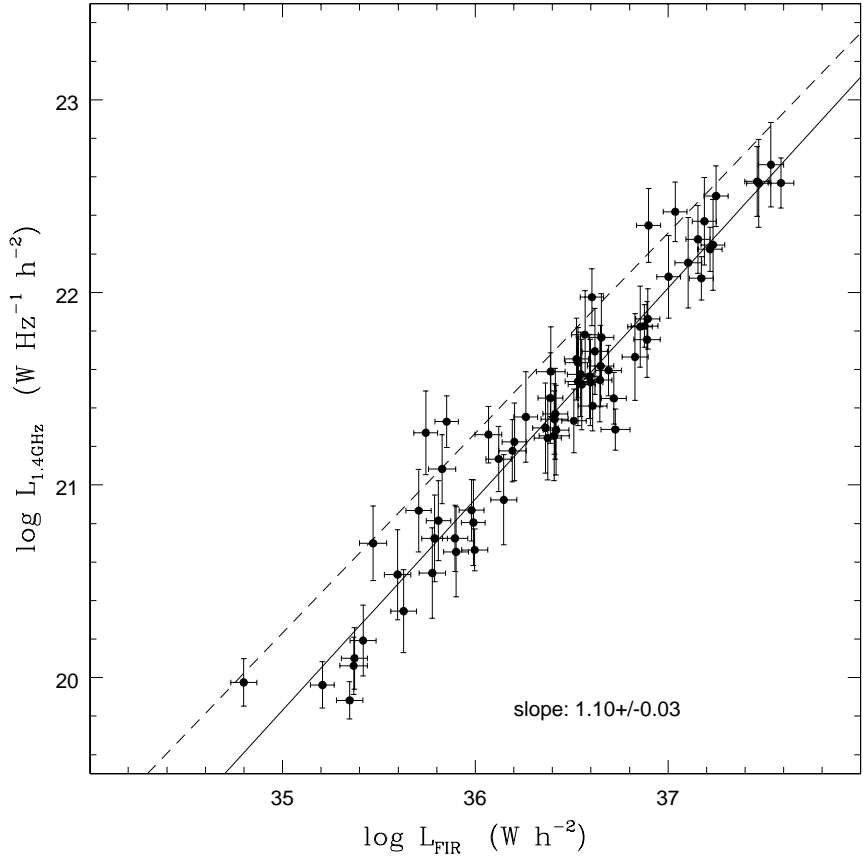}
      \caption{The radio luminosity at 1.4 GHz ($L_\mathrm{1.4 GHz}$)
               vs. the total FIR luminosity (of the warm $+$ cold dust
               emission components) ($L_{\mathrm{FIR}}$) for the full sample of
               72 objects under investigation. The plot shows a tight
               correlation between $L_{\mathrm{1.4 GHz}}$
               and $L_{\mathrm{FIR}}$, over three orders of magnitude
               in both luminosities. From the fit (solid line) we conclude that
               this correlation is only slightly non-linear (see the slope
               of the fit and its uncertainty, here indicated). The dashed line
               represents the fit to the $L_{\mathrm{1.4 GHz}}$
               vs. $L_{40-120}$ correlation, which is the equivalent of what
               would have been derived from IRAS.}
         \label{Fig1}
   \end{figure}

   Fig. 1 confirms that there is a tight correlation between
   dust and radio continuum emission, extending over three orders of magnitude
   in both luminosities.
   The total FIR/radio correlation is only slightly non-linear, with a slope
   $m=1.10 \pm 0.03$. We note that a luminosity-luminosity correlation analysis
   will be biased towards linearity, because it includes also the scaling
   effect due to galaxy size (or mass).
   Therefore, it is to be expected that the derived slope is only a lower limit
   to the slope that would be obtained from an analysis of
   normalised luminosities (see Xu et al. 1994). 

   As a comparison, in Fig. 1 we also plot the fit to
   the $L_\mathrm{1.4 GHz}$ vs. $L_{40-120}$ correlation, which is
   the equivalent of what would have been derived from IRAS.
   The statistics for this fit are also given in Table 3.
   There is an obvious shift of up to 0.4 in $\log L_{\mathrm{FIR}}$,
   which is produced by the inclusion of the cold dust component,
   mainly not seen by IRAS.
   This shift is more pronounced at the faint end of the correlation,
   where the cold dust component is more dominant.
   This also leads to a slight increase in non-linearity
   when the total FIR luminosity is taken into account instead of $L_{40-120}$.

   The fact that some galaxies show evidence for mild LINER/Seyfert activity
   or are in pair/multiple/merging systems does not seem to affect the shape
   of the correlation (not shown).
   In fact, we obtain $m=1.09 \pm 0.07$ for the subsample of 20 galaxies
   with some activity (Table 3) and $m=1.10 \pm 0.04$ for the remainder. 
   Since accretion powered systems are known not to lie on the FIR/radio
   correlation (Sopp \& Alexander 1991), this implies that even for the objects
   in our sample with some evidence for AGN activity, the FIR emission
   is powered by star formation activity and not by accretion
   (cf. Rodriguez Espinosa, Rudy \& Jones 1987).

\subsection{The warm FIR/radio correlation}
   \begin{figure}[htb]
   \centering
   \includegraphics[width=12cm]{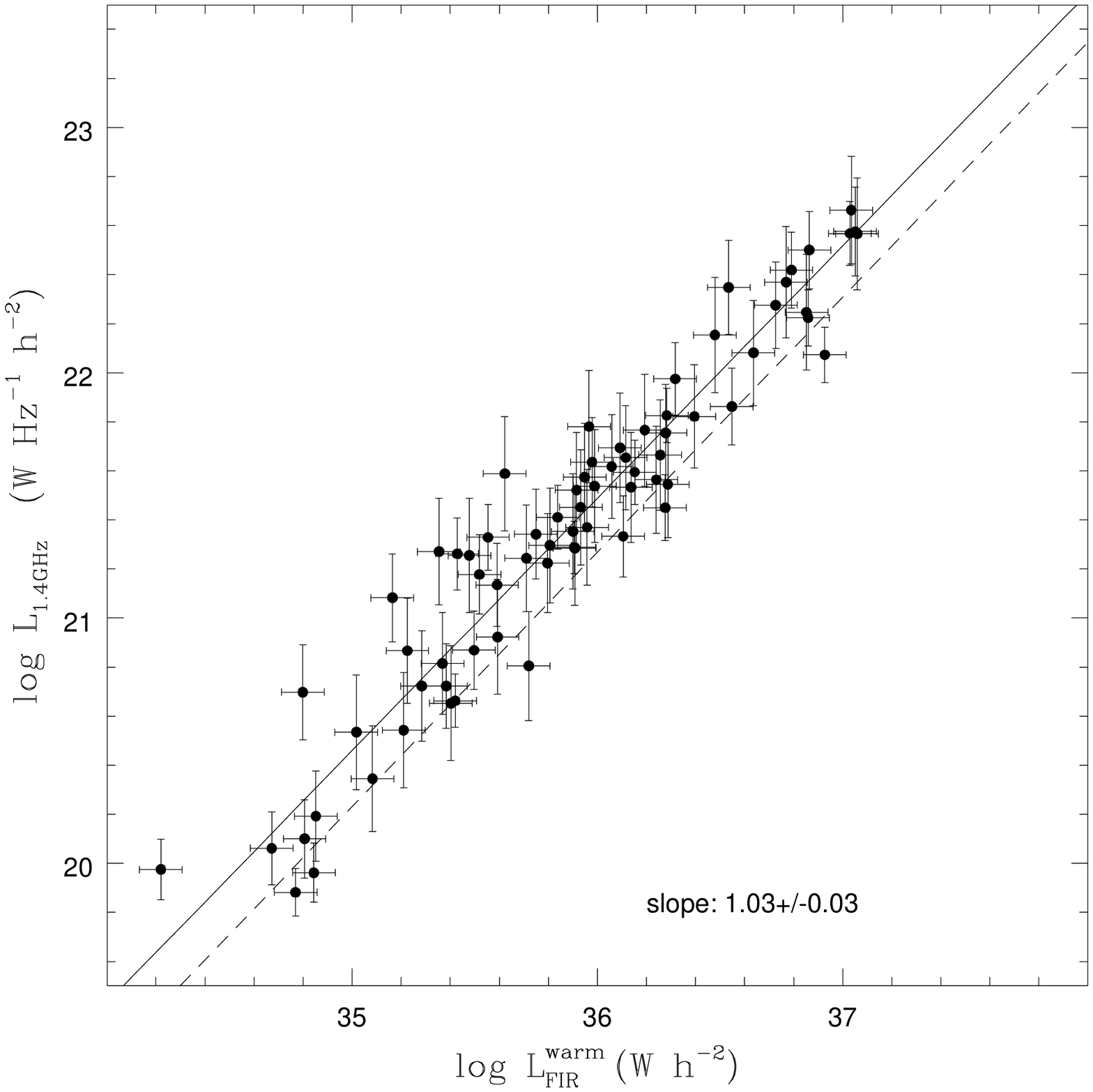}
      \caption{The radio luminosity at 1.4 GHz vs. the FIR luminosity
               of the warm component ($L_{\mathrm{FIR}}^{\mathrm{warm}}$)
               for the full sample of 72 objects under investigation.
               The plot shows a tight correlation between these two emissions,
               over three orders of magnitude in both luminosities.
               The fit to the correlation (plotted with the solid line)
               indicates a linear relation. The value of the slope
               and its uncertainty are indicated on the figure. The dashed line
               represents again the fit to the $L_{\mathrm{1.4 GHz}}$ vs.
               $L_{40-120}$ correlation, i.e., the equivalent of the IRAS
               FIR/radio correlation.}
         \label{Fig2}
   \end{figure}

   Fig. 2 reproduces the relationship between the warm FIR luminosity
   and $L_\mathrm{1.4 GHz}$ for the 72 sample galaxies.
   The slope of the correlation is $m=1.03 \pm 0.03$ (see Table 3),
   which indicates that the correlation is linear.

   This result differs from the previous one of Xu et al. (1994), who derived
   the warm FIR/radio correlation based on the proportionality between
   the IRAS FIR and H$\alpha$ emissions.
   These authors obtained a non-linear correlation, with the radio-to-FIR
   luminosity ratio increasing with luminosity.
   One possible explanation for the non-linearity of the correlation obtained
   by Xu et al. (1994) is that the H$\alpha$ emission was corrected for
   extinction using the same value for all galaxies.
   However, if more luminous galaxies suffer from larger amounts of
   extinction, then the intrinsic H$\alpha$ emission
   at the bright end of the correlation was underestimated by Xu et al.,
   and thus the warm FIR emission was also underestimated, producing
   the non-linearity of the warm FIR/radio correlation found by these authors.

   As for the total FIR/radio correlation, the fact that some galaxies show 
   evidence for mild LINER/Seyfert activity or are in pair/multiple/merging
   systems does not seem to affect the shape of the correlation (not shown).
   The slope of the correlation for the 20 galaxies with some activity
   is $m=0.96 \pm 0.06$ (Table 3).
   For the remainder we obtain $m=1.06 \pm 0.04$.

   The comparison between the warm FIR/radio correlation and the equivalent of
   the IRAS FIR/radio correlation (plotted with the dashed line in Fig. 2) 
   shows that the shift between the two correlations is much smaller than
   in the case of the total FIR luminosity.
   In addition, this shift is constant along the luminosity range
   spanned by the data.
   This is to be expected, as the IRAS $L_{40-120}$ luminosity is dominated by
   the contribution of the warm dust component, and contains only
   a small contribution from the cold dust component.

\subsection{The cold FIR/radio correlation}
   \begin{figure}[htb]
   \centering
   \includegraphics[width=12cm]{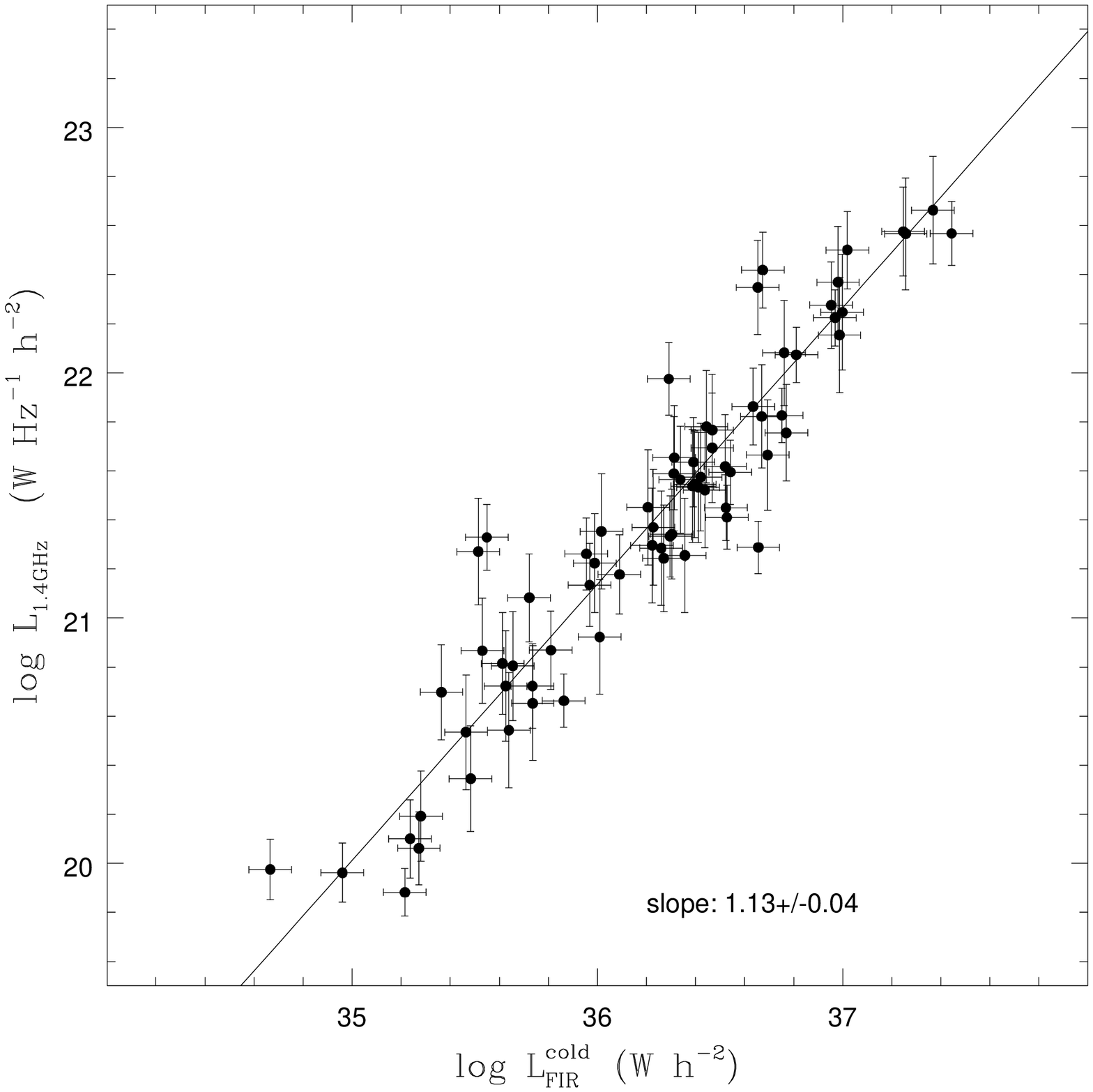}
      \caption{The radio luminosity at 1.4 GHz vs. the FIR luminosity
               of the cold component ($L_{\mathrm{FIR}}^{\mathrm{cold}}$)
               for the full sample of 72 objects under investigation.
               The plot shows a tight correlation between these two emissions,
               over three orders of magnitude in both luminosities.
               The fit to the correlation (plotted with solid line) indicates a
               slightly nonlinear correlation. The value of the slope and its
               uncertainty are indicated on the figure.}
         \label{Fig3}
   \end{figure}

   Fig. 3 reproduces the cold FIR/radio correlation as obtained
   for the whole sample of 72 objects.
   We find that the cold FIR/radio correlation is only slightly non-linear
   ($m=1.13 \pm 0.04$, see Table 3), similar with what we found for 
   the correlation between $L_{\mathrm{1.4 GHz}}$
   and the total FIR luminosity $L_{\mathrm{FIR}}$ (Fig. 1).
   This is no surprise since the cold dust component dominates
   the total FIR continuum emission of normal late-type galaxies.

   We obtained $m=1.17 \pm 0.08$ (Table 3) for the subsample of 20 galaxies 
   with some activity (LINER/Seyfert or pair/multiple/merging systems)
   and $m=1.12 \pm 0.05$ for the remainder.
   Thus, the different mix of galaxies does not seem to play
   any significant role, as for all correlations presented in this study.

\subsection{The FIR/radio correlation for the Virgo and non-Virgo galaxies}

   In their study of cold dust in Virgo Cluster galaxies, Popescu et al. (2002)
   plotted warm and cold FIR/radio correlations based on the ISO measurements
   for the 13 galaxies in their sample which had corresponding radio data.
   Although the statistics were poor, there was an indication that
   the warm FIR/radio correlation might be non-linear, whereas in this paper
   we obtain a linear slope for our sample of 72 objects.
   Since the sample presented in this work includes the Virgo galaxies
   from Popescu et al. (2002), we expect to reproduce the same trend
   when the fit is only applied to the Virgo galaxies.
   For this we fitted once again the FIR/radio correlations
   (for the total FIR emission, as well as for the warm and cold
   FIR emissions), this time separately for the Virgo and non-Virgo galaxies.
   In total there are 17 Virgo cluster galaxies, comprising the 14 from
   the VIRGO subsample and 3 Virgo cluster galaxies detected by ISS
   but not observed by Tuffs et al. (2002a,b).
   Correspondingly, there are 55 non-Virgo galaxies.

   The statistics of the fits are given in Table 3.
   Indeed, in the case of the warm FIR/radio correlation, there is
   some indication for a non-linear correlation,
   while the fit to the non-Virgo sample is linear.
   Overall, there is a tendency for all correlations to be more non-linear
   for the Virgo sample than for the non-Virgo sample.

   In this context it is appropriate to examine the question of
   environmental effects.
   Such influence on the FIR emission was not found for the Virgo galaxies
   (Popescu et al. 2002; Popescu \& Tuffs 2002b).
   However, the radio emission may be more susceptible to
   environmental effects, as the radio emitting medium is less tightly bound
   to the disk than the FIR emitting medium.
   One could imagine a scenario in which synchrotron halo electrons
   are swept away by the ram pressure induced by the relative motion of
   the galaxies through the intracluster medium.
   Preferentially, such a removal may be more severe for more quiescent
   galaxies which have less strongly magnetised halos.
   This could induce a non-linearity in the FIR/radio correlation.
   However this should only be applicable to galaxies which are entering
   the intracluster medium now and therefore we expect only sporadic examples
   of this phenomenon.
   We conclude that environmental effects play only a secondary role in shaping
   the FIR/radio correlation.
   The main distinction between Virgo and non-Virgo galaxies is rather
   that the former populate the faint end of the correlation,
   and are more quiescent.
\begin{table}
\caption{The linear best-fit parameters of the FIR/radio correlation$^{\dagger}$.}
\begin{tabular}{|c|c|c|c|c|c|}
\hline
                         &               & whole galaxy & galaxies with 
& Virgo & non-Virgo \\
                         &               & sample       & some activity 
& galaxies & galaxies\\
                         &               & N=72  & N=20 & N=17 & N=55 \\ 
\hline
                     & $m$            &   1.10 &   1.09 &   1.31 &   1.10 \\
                     & $\epsilon(m)$  &   0.03 &   0.07 &   0.12 &   0.06 \\
$L_\mathrm{1.4 GHz}$ vs. & $c$            & -18.53 & -18.35 & -26.29 & -18.68 \\
$L_{\mathrm{FIR}}$   & $\epsilon(c)$  &   1.23 &   2.52 &   4.23 &   2.07 \\
                     & $\chi^2_{\mathrm{red}}$ &   1.35 &   1.67 &   2.29 &   1.05 \\
                     & $\sigma$       &   0.13 &   0.14 &   0.15 &   0.12 \\

\hline
                     & $m$            &   1.03 &   0.96 &   1.27 &   0.98 \\
                     & $\epsilon(m)$  &   0.03 &   0.06 &   0.12 &   0.05 \\
$L_\mathrm{1.4 GHz}$ vs. & $c$            & -15.55 & -12.80 & -23.68 & -13.88 \\
$L_{\mathrm{FIR}}^{\mathrm{warm}}$ & $\epsilon(c)$  &   1.19 &   2.22 &   4.23 &   1.88 \\
                     & $\chi^2_{\mathrm{red}}$ &   1.15 &   1.26 &   2.15 &   0.82 \\
                     & $\sigma$       &   0.14 &   0.15 &   0.17 &   0.13 \\
\hline
                     & $m$            &   1.13 &   1.17 &   1.35 &   1.14 \\
                     & $\epsilon(m)$  &   0.04 &   0.08 &   0.14 &   0.06 \\
$L_\mathrm{1.4 GHz}$ vs. & $c$            & -19.43 & -21.85 & -27.03 & -19.93 \\
$L_{\mathrm{FIR}}^{\mathrm{cold}}$ & $\epsilon(c)$  &   1.40 &   2.91 &   4.74 &   2.07 \\
                     & $\chi^2_{\mathrm{red}}$ &   1.61 &   2.27 &   2.36 &   1.38 \\
                     & $\sigma$       &   0.15 &   0.17 &   0.16 &   0.15 \\
\hline
                     & $m$            &   1.04 &   0.93 &   1.29 &   1.01 \\
                     & $\epsilon(m)$  &   0.03 &   0.06 &   0.10 &   0.05 \\ 
$L_\mathrm{1.4 GHz}$ vs. & $c$            & -16.13 & -12.24 & -25.02 & -15.11 \\
$L_{40-120}$         & $\epsilon(c)$  &   1.12 &   2.04 &   3.65 &   1.86 \\
                     & $\chi^2_{\mathrm{red}}$ &   1.28 &   1.61 &   2.85 &   0.75 \\
                     & $\sigma$       &   0.13 &   0.15 &   0.16 &   0.11 \\
\hline
\end{tabular}
\begin{list}{}{}
\item[$^{\dagger}$] in log--log
\end{list}
\end{table}
%

\section{Interpretation of the warm and cold FIR/radio correlations}

   In the previous section we have, for the first time, quantitatively
   established the correlation between the FIR and radio emission by including
   the bulk of the FIR luminosity carried by the cold dust component.
   Furthermore our measurements at three FIR wavelengths allowed
   a direct determination of the warm and cold FIR emissions.
   Consequently, separate relations have been derived for the warm
   and cold FIR/radio relations.
   In this section we will discuss each of the correlations in terms of
   the dependence on the star formation rate (SFR). 

   Detailed mapping observations in the 40-200\,${\mu}$m range of
   nearby galaxies (e.g. Haas et al. 1998; Hippelein et al. 2002) as well as
   self consistent models of the UV/submm SEDs in spiral galaxies
   (Popescu et al. 2000) are both consistent with the presence of
   locally heated dust in star-forming complexes and of diffuse dust
   (Tuffs \& Popescu 2002).
   The diffuse dust is heated both by non-ionising UV photons
   and optical photons (Xu 1990).
   Then, the relation between SFR and the total FIR energy output
   can be derived from the following equation:
\begin{eqnarray}
L_{\rm FIR}^{\rm tot}=L_{\rm FIR}^{\rm HII}+L_{\rm FIR}^{\rm UV}+L_{\rm FIR}^{\rm opt}
\end{eqnarray}
   where $L_{\rm FIR}^{\rm tot}$ is the total FIR luminosity emitted by
   the galaxy, $L_{\rm FIR}^{\rm HII}$ is the FIR luminosity emitted by
   the HII regions, $L_{\rm FIR}^{\rm UV}$ is the component of
   the diffuse FIR luminosity powered by the non-ionising UV photons
   and $L_{\rm FIR}^{\rm opt}$ is the component of the diffuse FIR luminosity
   powered by the optical photons.
   The equation can be further expressed\footnote{The formulation in Eq. (2)
   is applicable for the case that the diffuse component is illuminated by
   a continuous stellar emissivity, and neglects the discrete character of
   stars.} in terms of SFR:
\begin{eqnarray}
L_{\rm FIR}^{\rm tot}= SFR \times (L_0 \times F + L_x \times X) + 
SFR \times L_0 \times (1-F) \times G_{uv}+L_{\rm FIR}^{\rm opt}
\end{eqnarray}
   where the first term corresponds to $L_{\rm FIR}^{\rm HII}$
   and the second term to $L_{\rm FIR}^{\rm UV}$.
   SFR - the variable of Eq. 2 - is the present-day star formation rate
   in ${\rm M}_{\sun}/{\rm yr}$, $L_0$ and $L_x$ are the non-ionising
   and the ionising UV bolometric luminosities of a young stellar population
   corresponding to $SFR = 1\,{\rm M}_{\sun}$/yr (which can be derived from
   population synthesis models), F and X are the fractions of non-ionising
   and ionising UV emission that are absorbed by dust locally
   within star forming complexes, and the factor $G_{uv}$ is the probability
   that a non-ionising UV photon escaping from the star formation complexes
   will be absorbed by dust in the diffuse interstellar medium.   

   In the way it is defined (see also Popescu et al. 2000), the $F$ factor
   accounts for the inhomogeneities in the distributions of dust and stars.
   It determines the additional likelyhood of absorption of
   non-ionising UV photons due to correlations between an inhomogeneous
   distribution of young stars and (opaque) parent molecular clouds.
   In this sense the $F$ factor represents a clumpiness factor in galaxies. 

   The value of $G_{uv}$ depends on both the overall amount of diffuse dust
   and its geometrical distribution relative to the distribution of
   the young stellar population. 
   The relative distribution of stellar emissivity and dust as a function of 
   SFR is not known.
   But broadly speaking one would expect SFR to increase with the gas surface
   density in galaxy disks (Schmidt 1959; Kennicutt 1998).
   If the dust content is proportional to the gas content\footnote{Broadly
   speaking, the dust content is proportional to the metal abundance
   (e.g., [O/H]) times the average gas column density of the disk.
   [O/H] is known to increase with the optical luminosity of the galaxy
   (Zaritsky, Kennicutt \& Huchra 1994), enforcing the idea that
   the metal content of a galaxy like our own is mostly the result of
   the past star formation activity (e.g. Dwek 1998).
   In absence of an observationally established relation between
   metallicity of the disk and SFR, we expect this metallicity
   to contribute only to the scatter of the relation between dust content
   and SFR.}, the dust opacity would also increase with SFR
   (for a fixed geometry).
   $G_{uv}$ would then increase with SFR for the optically thin case
   and would tend asymptotically towards the fixed value of unity
   for the optically thick case.
   From this definition it is obvious that the $G_{uv}$ is a measure of opacity
   in galaxies.  

   The fraction X of ionising UV photons that are absorbed by dust
   in HII regions exhibits a broad range of values, varying from $0.3-0.7$
   (Inoue  et al. 2001, Inoue 2001).
   However, even if X approaches unity, the intrinsic luminosity $L_{\rm X}$
   is so much smaller than the intrinsic non-ionising UV luminosity $L_0$
   (Bruzual \& Charlot 1993), that its contribution to $L_{\rm FIR}^{\rm HII}$
   can still be neglected.

\subsection{The warm FIR/radio correlation}

   To interpret the warm FIR/radio correlation, we identify the warm dust
   component from our fitting procedure with dust locally heated within
   the HII regions, such that $L_{\rm FIR}^{\rm warm}\simeq L_{\rm FIR}^{\rm HII}$.
   In this case $L_{\rm FIR}^{\rm warm}$ can be written as:
\begin{eqnarray}
L_{\rm FIR}^{\rm warm}\simeq SFR\times L_0\times F
\end{eqnarray}

   In principle Eq. 3 can be used to constrain the dependence of
   the radio luminosity on SFR.
   In Sect. 3.2 we found a linear correlation between the warm FIR
   and radio luminosities for the sample as a whole.
   Under the assumption that the clumpiness factor $F$ does not vary with SFR,
   this would imply a linear relation between radio luminosity and SFR.
   This is in accordance with the prediction of the calorimeter theory
   (V\"olk 1989).

\subsection{The cold FIR/radio correlation}

   If the cold dust component is identified with diffuse dust heated by
   the interstellar radiation field, then:
\begin{eqnarray}
L_{\rm FIR}^{\rm cold} \simeq SFR \times L_0 \times (1-F) \times G_{uv} + L_{\rm FIR}^{\rm opt}
\end{eqnarray}

   To interpret Eq. (4) we have to take into account that the opacity
   ($G_{uv}$ factor) could vary systematically with SFR (see also Bell 2003).  
   As argued above, $G_{uv}$ should increase with SFR for the optically
   thin case and should approach unity for the optically thick case.
   This means that $L_{\rm FIR}^{\rm cold}$ would increase more 
   than linearly with increasing SFR for the optically thin case
   and would tend towards linearity for the extreme optically thick case. 
   In Sect. 3.3 we found a slightly non-linear  cold FIR/radio correlation 
   ($m=1.13$ with a $3\sigma$ lower limit on the slope of 1.01) for the sample
   as a whole, in the sense that the ratio radio-to-FIR increases more than 
   linearly with increasing luminosity, the opposite trend than that predicted
   by opacity.
   Could this observed trend be explained at low luminosities by 
   considering an increased contribution of the heating of dust by
   optical photons?
   Xu et al. (1994) studied the effect of the optical heating on
   the FIR/radio correlation and found that this would indeed introduce
   a non-linearity at low luminosities.
   However they assumed that $G_{\rm uv}$ was the same for all galaxies,
   irrespective of SFR.
   In fact the opacity should have its strongest dependence on SFR
   at low luminosities.
   As argued by Bell (2003), the effect of optical heating is negligible 
   compared with the effect of opacity.
   Thus, most probably, the optical heating is not enough to explain
   the observed slope at low luminosities.
   One possibility to explain the observed slope of the cold FIR/radio 
   correlation is that the radio emission increases more than linearly
   with increasing SFR for the optically thin case ($G_{uv} <1$)
   and has a marginal deviation from linearity for the optically thick case
   ($G_{uv}$ approaches 1).
   But this is in contradiction with our inference from the warm FIR/radio
   correlation that the radio luminosity has a linear dependence on SFR.
%

\section{Discussion}

   We have seen in the previous section that it is not easy to identify
   a scenario which can simultaneously account for both the warm and cold
   FIR/radio correlations.
   Whereas the warm FIR/radio correlation was consistent with
   a linear dependence of the radio on SFR, the cold FIR/radio correlation
   was more readily interpreted in terms of a non-linear dependence of
   the radio emission on SFR.
   To reconcile these seemingly conflicting results we need to re-examine
   the underlying assumptions on which the results were based.

   In drawing the conclusion from the warm FIR/radio correlation we assumed
   that the clumpiness factor ($F$) is a constant.
   However, if an increased SFR is accompanyed not only by an increase
   in the number of independent HII regions, but also by a higher probability
   for further star formation to happen preferentially near already existing
   HII regions, then the $F$ factor would also increase, as a consequence of
   the increased blocking capability of the optically thick molecular clouds
   in the star-forming complex (e.g. Witt \& Gordon 1996).
   This would be expected to occur if star formation is a self propagating
   phenomenon, in which preceding generations of stars can trigger
   the formation of new generations.  

   Referring again to Eq. 3, we see that, if the clumpiness factor $F$
   increases with SFR, then the $L_{\rm FIR}^{\rm warm}$ increases more rapidly
   than linearly with SFR.
   In turn, the radio luminosity, which empirically was found to be
   proportional to $L_{\rm FIR}^{\rm warm}$, will also increase more rapidly
   than linearly.
   Qualitatively this is what is needed to bring the warm FIR/radio correlation
   into consistency with the cold FIR/radio correlation.
   In Eq. 4 the effect of increasing the clumpiness factor $F$ with SFR
   would depress $L_{\rm FIR}^{\rm cold}$.
   This, together with the non-linear dependence of the radio on the SFR,
   would mitigate the trends introduced by opacity, leading to an increased
   ratio radio/cold FIR with increasing SFR.  
 
   From this discussion we conclude that one possibility to bring
   into consistency the warm and cold FIR/radio correlations would be to invoke
   a non-linear dependence of the radio emission on SFR.
   Such a non-linear dependence could be understood in terms of an increase
   in the residence time of the synchrotron-emitting electrons with increasing
   galaxy size, as postulated by Chi \& Wolfendale (1990).
   However, the same trend can also be obtained within the framework of
   the calorimeter theory, if the radio emission is identified with the sum of
   diffuse and localised source components.
   The localised source component is comprised of synchrotron emission
   from supernova remnants (which account for about $10\%$ of the total
   radio flux density of galaxies at 1.4\,GHz, Lisenfeld \& V\"olk 2000)
   plus thermal emission from star forming regions.
   The diffuse component consists of a pure synchrotron emission component,
   as given by V\"olk (1989), plus an additional diffuse free-free component.
   The free-free component arises from the warm ionised medium, as defined by
   McKee \& Ostriker (1977), and its emission is in general much smaller than
   the synchrotron emission.
   Nevertheless we expect that the relative contribution of the synchrotron
   and free-free components to vary with SFR, such that at high SFR
   the free-free component becomes non-negligible.
   This is because the filling factor of the warm ionised medium increases
   with increasing SFR.
   Since the non-thermal diffuse radio emission is still proportional with SFR
   (V\"olk 1989), then the increased fraction of free-free emission
   with increasing SFR will introduce the non-linearity in the radio emission. 

   In this paper we have used a physical though qualitative description of
   the warm and cold FIR/radio correlations.
   To quantitatively model these correlations we need to derive the dependence
   on SFR of both the FIR and the radio emissions.
   From the point of view of the FIR emission, we need to self consistently
   analyse the UV to submm SED of each individual galaxy (Popescu et al. 2000;
   Popescu \& Tuffs 2002).
   In this way the variation of the terms in Eqns. 3 \& 4 - clumpiness
   (the F factor) and opacity (G factor) - with SFR can be derived.
   From the point of view of the radio emission we need to determine
   the relative contribution of its components and express them in terms of
   the SFR.
   Such a quantitative treatment would throw insights into
   the physical processes that make normal galaxies the ``machines'' they are,
   ranging from particle accelerators and magnetic field dynamos,
   to radiative coolants.

\section{Summary}

   In this paper we present the first analysis of the FIR/radio correlation
   which incorporates the bulk of the FIR luminosity radiating longwards of
   the spectral coverage of IRAS.
   For this we constructed a composite sample of normal galaxies measured with
   the ISOPHOT instrument on board ISO.
   The sample was a combination of the ISOPHOT Virgo Cluster Deep Sample
   (Tuffs et al. 2002a,b) and of the ISOPHOT Serendipity Sample
   (Stickel et al. 2000).
   This provided us with a high dynamic range in star formation activity
   as well as good statistics.
   All galaxies in the composite sample have had their integrated flux
   densities measured by ISOPHOT at 170\,${\mu}$m and by ISOPHOT (Virgo sample)
   or IRAS (Serendipity sample) at 60 and 100\,${\mu}$m.
   The luminosities of the warm and cold dust emission components
   were extracted from the available measurements at the three FIR wavelengths.
   This allowed us to define the FIR/radio correlation separately for
   the warm and cold dust components.

   The inclusion of the cold dust component produces a tendency
   for the total FIR/radio correlation to become non-linear.
   We found that the cold FIR/radio correlation is slightly non-linear,
   whereas the warm FIR/radio correlation is linear.
   In order to interpret the individual warm and cold correlations in terms of
   SFR, we identified the warm emission component with dust locally
   heated within the HII regions, whereas the cold emission component
   was identified with diffuse dust heated by the interstellar radiation field.
   Because the effect of opacity in galaxies would introduce a non-linearity
   in the cold-FIR/radio correlation, in the opposite sense to that observed,
   both the radio and the FIR emissions are likely to have
   a non-linear dependence on SFR (for the range of luminosities covered by
   our sample, which lies below the luminosity range if starburst galaxies).
   The behaviour of the radio emission can be accounted for by considering
   the radio emission to be a superposition of a diffuse synchrotron component,
   a diffuse free-free component and a localised source component. 
%

\begin{acknowledgements}
      This research has made use of the NASA/IPAC Extragalactic Database
      (NED), which is operated by the Jet Propulsion Laboratory, California
      Institute of Technology, under contract with the National Aeronautics
      and Space Administration.
\end{acknowledgements}

%

\appendix

\section{The extraction of the radio flux densities from the NVSS maps}

   The NVSS used the NRAO Very Large Array telescope and covers the sky
   north of a declination of -40 degrees at a frequency of 1.4 GHz,
   a resolution of $\rm 45^{\prime \prime}$ and a limiting peak source
   brightness of about 2.5 $\rm mJy~beam^{-1}$ (Condon et al. 1998).
   Radio counterparts are associated with optically catalogued galaxies
   when the distance of the peak radio surface brightness of
   the potential radio identification from the optical position of
   the target galaxy is within the NVSS resolution.
   The probability of finding an NVSS source within $\rm 45^{\prime \prime}$
   of an arbitrary position is 0.02.
   \begin{figure}[htb]
   \centering
   \includegraphics[width=12cm]{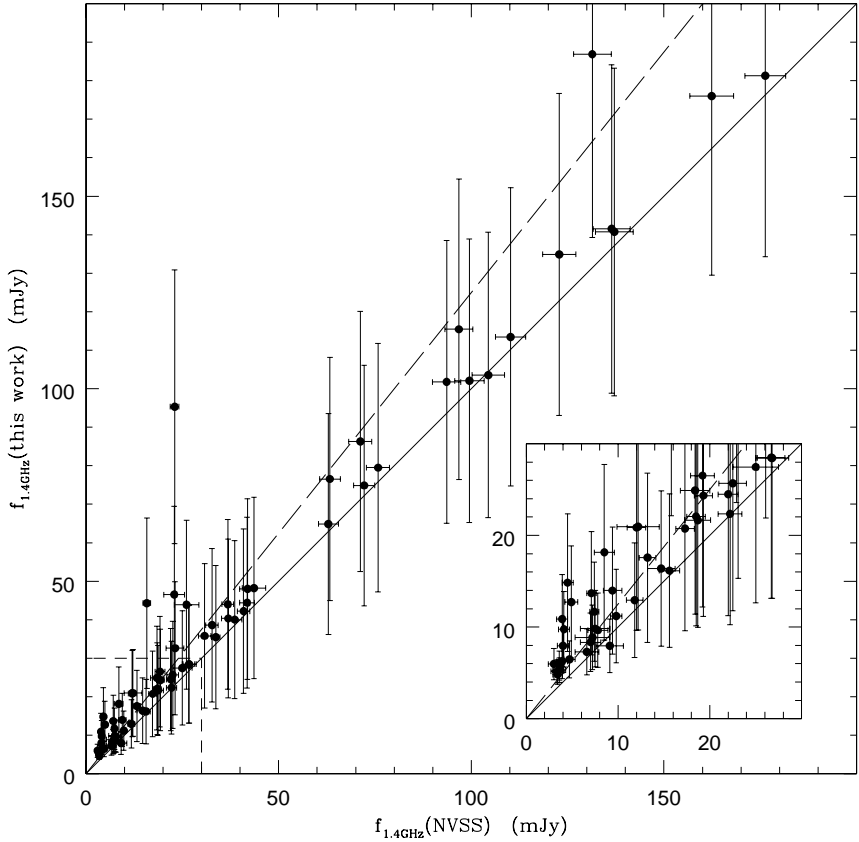}
      \caption{Comparison of the radio fluxes at 1.4 GHz determined in this
               study from direct aperture photometry of the NVSS images and
               the NVSS catalogue fluxes for the radio counterparts of
               the 72 sample objects under investigation. The latter fluxes
               are obtained from Gaussian model components fitted to
               the survey images. Here the solid and long-dashed lines
               represent the 1:1 and 1.25:1 ratios, respectively, while
               the box delimited by a short-dashed line corresponds to
               the region zoomed in the enclosed box. On average, our fluxes
               are 25 per cent larger than the NVSS catalogue one (see text).}
         \label{FigA1}
   \end{figure}

   As a result of this survey a source catalogue has been produced,
   which consists of Gaussian model components fitted to the survey images.
   If the source is significantly larger than the resolution of the survey
   then it may be represented by several of these components.
   The fitted parameters are deconvolved from the instrumental resolution
   producing source sizes (or upper limits with 99 per cent confidence)
   and the integrated flux density of the model.
   Error estimates for each of the source parameters are also given.

   In addition to the catalogue, a postage stamp server which returns
   FITS files of limited regions from the NVSS calibrated images
   is available electronically.
   The noise level in these images is about 0.5 $\rm mJy~beam^{-1}$. 
   We have decided to extract our own aperture photometry from
   these calibrated maps under the consideration that some
   of our sample galaxies might have failed the classification criteria
   of the NVSS catalogue and/or not be represented by Gaussian model components
   at best and/or have unaccounted faint extended emission.
   This question is crucial, since the $\rm 45^{\prime \prime}$ beam
   of the NVSS has a much smaller aperture than the ISOPHOT beam
   at $\rm 170~\mu m$ ($\rm 92^{\prime \prime}$).
   Aperture photometry was extracted with the task QPHOT
   in IRAF\footnote{IRAF is the Image Analysis and Reduction Facility
   made available to the astronomical community by the National Optical
   Astronomy Observatories, which are operated by AURA, Inc., under contract
   with the U.S. National Science Foundation.}.

   We plot the comparison between our photometry (y axis)
   and the NVSS catalogue one (x axis) in Fig. A.1.
   Here the solid and long-dashed lines represent the 1:1 and 1.25:1 ratios,
   respectively, while the box delimited by a short-dashed line corresponds
   to the region zoomed in the enclosed box.
   As expected, we find that the agreement is excellent for those radio sources
   which appear unresolved with the NVSS beam but it becomes worse in case of
   very extended/complex sources.
   On average, our fluxes are 25 per cent larger than the NVSS catalogue one.
%

\section{Photometric comparison between the ISOPHOT Serendipity Survey and 
the ISOPHOT Deep Virgo Cluster Survey at 170\,$\mu m$}

   Four galaxies from the ISOPHOT Deep Virgo Cluster Survey (Tuffs et al. 2002)
   were also detected by the ISOPHOT $\rm 170~\mu m$ Serendipity Survey
   (Stickel et al. 2000).
   These galaxies are: VCC\,66, 460, 836 and 873, alias NGC\,4178, 4293,
   4388 and 4402, respectively.
   From the comparison of the $\rm 170~\mu m$ flux densities of
   these 4 galaxies it emerges that the values of Stickel et al.
   are systematically higher than those of Tuffs et al.,
   the difference being significant (a factor of 2) for VCC\, 836 and 873.

   The absolute calibration is not a reason for this,
   since background measurements averaged over the 4 common objects
   are the same to within a few per cent.
   However the different techniques used to derived photometry may give rise
   to differences.
   Even for extended sources, the Serendipity Survey photometry is extracted
   by fitting circular gaussians to the strip maps, whereas the Virgo sample
   photometry is performed by fitting elliptical gaussians.
   Thus, one may expect the Serendipity Survey $\rm 170~\mu m$ fluxes
   to be overestimated as inclination gets larger, as indeed it is the case
   for the previous 4 galaxies.
   In addition, in the Serendipity Survey photometry there is a spread
   of 30 per cent for bright sources, due to systematic effects like offset
   of source from scan line.
   To this must be added random errors for faint sources (10 Jy or less).
   The combined uncertainties can be 50 per cent for faint sources
   (see Fig. 2 of Stickel et al. 2000).
   Finally, compared with Tuffs et al. (2002), Stickel et al.
   made larger corrections for the loss of signal incurred by
   the transient response of the detector to illumination steps --
   the so-called ``transient correction''.
   The multiplicative corrections of Stickel et al. range from 1.1
   for a faint source to a factor of 2 for sources of 60 Jy.
   By comparison the transient corrections of Tuffs et al.
   are 10 to 20 per cent only for the C200 detector.

   From this discussion we conclude that the Serendipity Survey photometry
   may be overestimated by a factor of 2 for large (i.e., with apparent
   major axis longer than $\rm \sim 4^{\prime}$) galaxies with inclinations
   higher than $\rm \sim 70^o$.

\end{document}